\documentclass[sigconf,authorversion]{acmart}

\usepackage{booktabs}  
\usepackage{pbox}
\usepackage{subfigure}
\usepackage{epsfig,endnotes}
\usepackage{epstopdf}
\usepackage{times}
\usepackage{color}
\usepackage{graphicx}
\usepackage{xspace}
\usepackage{flushend}  
\usepackage{enumitem}
\usepackage{url}

\newif\ifcomment
\commenttrue

\ifcomment
    \newcounter{MVNumberOfComments}
    \stepcounter{MVNumberOfComments}
    \newcommand{\mvnote}[1]{\textcolor{blue}{\small \bf [MV\#\arabic{MVNumberOfComments}\stepcounter{MVNumberOfComments}: #1]}}
    \newcounter{DPNumberOfComments}
    \stepcounter{DPNumberOfComments}
    \newcommand{\dpnote}[1]{\textcolor{green}{\small \bf [DP\#\arabic{DPNumberOfComments}\stepcounter{DPNumberOfComments}: #1]}}
    \newcounter{CSNumberOfComments}
    \stepcounter{CSNumberOfComments}
    \newcommand{\csnote}[1]{\textcolor{red}{\small \bf [CS\#\arabic{CSNumberOfComments}\stepcounter{CSNumberOfComments}: #1]}}

    \newcommand{\NOTE}[1]
    {
      {\footnotesize\it
        \begin{center}
          \begin{tabular}{|c|}
           \hline
            \parbox{0.85\columnwidth}{
              \medskip
              #1
              \medskip} \\
            \hline
          \end{tabular}
        \end{center}
        }
    }
\else
    \newcommand\mvnote[1]{}
    \newcommand\dpnote[1]{}
    \newcommand\csnote[1]{}
    \newcommand\NOTE[1]{}
\fi

\setcopyright{rightsretained}

\newcommand{\eg}{{e.g.,}\xspace}
\newcommand{\ie}{{\it i.e.,}\xspace}

\newcommand\co[1]{}

\newcommand{\tool}{{\sf{ProxyTorrent}}\xspace}
\newcommand{\plugin}{{\sf{Ciao}}\xspace}

\smallskip
\smallskip

\newcommand{\takeaway}[1]{\noindent \textit{\textbf{Takeaway:}~#1}}

\begin{document}

\title{ProxyTorrent: Untangling the Free HTTP(S) Proxy Ecosystem}

\author{Diego Perino}
\affiliation{%
  \institution{Telefonica Research}
}
\email{diego.perino@telefonica.com}

\author{Matteo Varvello}
\authornote{Work done while at Telefonica Research}
\affiliation{%
  \institution{AT\&T Labs - Research}
}
\email{varvello@research.att.com}

\author{Claudio Soriente}
\affiliation{%
  \institution{NEC Laboratories Europe}
}
\email{claudio.soriente@emea.nec.com}

\begin{abstract}
Free web proxies promise anonymity and censorship circumvention at no cost. Several websites publish lists of free proxies organized by country, anonymity level, and performance. These lists index hundreds of thousand of hosts discovered via automated tools and crowd-sourcing. A complex free proxy \emph{ecosystem} has been forming over the years, of which very little is known. In this paper we shed light on this ecosystem via \tool, a distributed measurement platform that leverages both \emph{active} and \emph{passive} measurements. Active measurements discover free proxies, assess their performance, and detect potential malicious activities. Passive measurements relate to proxy performance and usage in the wild, and are collected by free proxies users via a Chrome plugin we developed.  \tool has been running since January 2017, monitoring up to 180,000 free proxies and totaling more than 1,500 users. Our analysis shows that less than 2\% of the proxies announced on the web indeed proxy traffic on behalf of users; further, only half of these proxies have decent performance and can be used reliably.  Around 10\% of the working proxies exhibit malicious behaviors, \eg ads injection and TLS interception, and these proxies are also the ones providing the best performance. Through the analysis of more than 2~Terabytes of proxied traffic, we show that web browsing is the primary user activity. Geo-blocking avoidance is not a prominent use-case, with the exception of countries hosting popular geo-blocked content.

\end{abstract}
\maketitle

\section{Introduction}
\label{sec:intro}
Web proxies are intermediary boxes enabling HTTP (sometimes also HTTPS) connections between a client and a server. They are widely used for security, privacy, performance optimization or policy enforcement, to cite a few use cases. Many web proxies are free of charge and publicly available. Such proxies can be used, for example, for private web surfing and to access content that would be blocked otherwise (\eg due to geographical restrictions).

Specialized forums, websites, and even VPN service providers\footnote{For example, \url{https://hide.me}} compile daily lists of free web proxies. When tested, most of these proxies are slow, unreachable or not even real proxies. Furthermore, it is folklore that free web proxies perform malicious activities, \eg injection of advertisements and user fingerprinting. It is fair to say that free proxies form a massive and complex \emph{ecosystem} of which very little is known. For example, what is the magnitude of the ecosystem and how many proxies are safe to use? How and for what are these proxies used? Answering these questions is hard
because of the sale of the ecosystem  and because it involves two players out of reach: free proxies and their users.


In this work we tackle the above challenge by building \tool, a distributed measurement platform of the free proxy ecosystem. \tool leverages our premises to \emph{actively} discover and assess the performance of the core of the ecosystem that can be used safely. Usage statistics are instead \emph{passively} (and anonymously) collected at free proxy users in exchange of high-quality proxies list compiled by \tool.

\tool is daily fed with tens of thousands \emph{potential} proxies obtained by crawling the most popular free proxies aggregator websites. Potential proxies are quickly tested in order to discard the ones that are unreachable, do not proxy traffic, or perform malicious activities. This is done by loading a ``bait'' webpage we have crafted as well as few popular webpages, and comparing the content received via the proxy with the one received when no proxy was set. The same approach is used to detect issues with X.509 certificates in case of TLS connections. These operations run daily at our premises and generate a few thousand \emph{trusted} proxies. Next, we test the performance of trusted proxies from $\sim30$ network locations (Planetlab nodes~\cite{planetlab}) while fetching the landing pages of popular websites via both HTTP and HTTPS. Collected data is finally used to populate a list of \emph{good} proxies, \ie working and trustworthy free proxies.

This list of good proxies is then offered to a Chrome plugin (\plugin~\cite{opencode, chrome_plugin}) we developed to help users interacting with the free proxy ecosystem. \plugin users select a target anonymity level and country, and the plugin automatically identifies the best free proxy for the task, if any. As the user browse the Internet through the proxy, we collect anonymous statistics on free proxy performance and how they are used in the wild.


We use data collected by \tool to provide a unique overview of the free proxy ecosystem. In this paper we present ten months worth of data spanning up to 180,000 free web proxies and more than 1,500 users. The analysis of this data-set reveals the following key findings:


\vspace{-0.1in}
\hfill \break
\noindent{\bf The free proxy ecosystem is large and ever-growing, but only a small fraction of the announced proxies actually works.} While thousands of new free proxies are announced daily, overall, less than 2\% of them are reachable and correctly proxy traffic. Further, half of these proxies stop working after few days. Many reasons are behind such ephemeral behavior: host misconfigurations, dynamic addressing, and even bait proxies from VPN providers aiming at attracting more customers.

\vspace{-0.1in}
\hfill \break
\noindent{\bf A non negligible percentage of working proxies are suspicious, but provide better performance than safe proxies.} Every day, around 10\% of the working proxies announced on the Web exhibit suspicious behavior, from injection of advertisements to TLS man-in-the-middle attempts. On average, these proxies are twice as fast as non-malicious ones. Fast connectivity is likely used to attract potential ``victims'', supporting the general belief that free proxies are ``free for a reason''.

\vspace{-0.1in}
\hfill \break
\noindent{\bf The geographical distribution of proxies is fairly skewed.} Half of the working free proxies reside in a handful of countries, with US, France, and China at the top. While US and China are at the top due to their sizes, the presence of large cloud providers in France is the reason behind such large number of proxies.

\vspace{-0.1in}
\hfill \break
\noindent{\bf Geo-blocking avoidance is not a prominent use-case for free web proxies.} By analyzing 2 TBytes of traffic generated by 1,500 \plugin users over 7 months, we conclude that web browsing is the most prominent activity. Proxy are rarely selected in the same location where a visited  website resides, which suggests that circumvention of potential geo-blocking rules is not a primary user concern. Some countries like the US are an exception though, likely because hosting a lot of popular geo-blocked content.



\section{Background and Related work}
\label{sec:back-related}
\vspace{-0.1in}
\hfill \break
\noindent\textbf{Background} -- A \emph{web proxy} is a device/application that acts as an intermediary for HTTP(S) requests, such as \texttt{GET} and \texttt{CONNECT}, issued by clients seeking resources on servers. Web proxies are commonly classified as \emph{transparent}, \emph{anonymous}, and \emph{elite},  depending on the degree of anonymity they provide.

Transparent proxies reveal the IP address of the client to the origin server, \eg by adding the \texttt{X-FORWARDED-FOR} header which specifies the address of the client. Anonymous proxies block headers that may allow the origin server to detect the identity of the client, but still announce themselves as proxies, \eg by adding the \texttt{HTTP\_VIA} header. Elite proxies do not send any of the above headers and look just like regular clients to the origin server. Yet, the origin server may detect that a proxy is being used by probing the IP address extracted from the received traffic to check if it acts as a proxy.



\hfill \break
\noindent\textbf{Related Work} --  We briefly overview relevant results from related work and highlight differences with \tool.

\noindent\emph{Free Web Proxies.}
Scott et al.,~\cite{scott15ccc} also study free web proxies but both their goal and methodology differ from ours. While our goal is a complete view of the free proxy ecosystem, they mostly focus on how and for what free proxies are used. They do so by scanning the IPv4 address space at popular proxy ports (\eg 3128, 8080, and 8123) looking for open management interfaces (\ie proxy interfaces with no authentication required) from which they can ``steal'' usage statistics. This approach is intended to run seldomly due to the cost associated with IPv4 scanning. Further, it raises some ethical concerns related to exposing hosts found via scanning as well as intruding their management interfaces. \tool was instead designed with both scalability and user privacy in mind. Note that we once ran a full scan of the IPv4 address space at popular proxy ports to compare with our methodology (see Table~\ref{tab:summ_magn}).

\begin{figure}[t]
\centering
	\psfig{figure=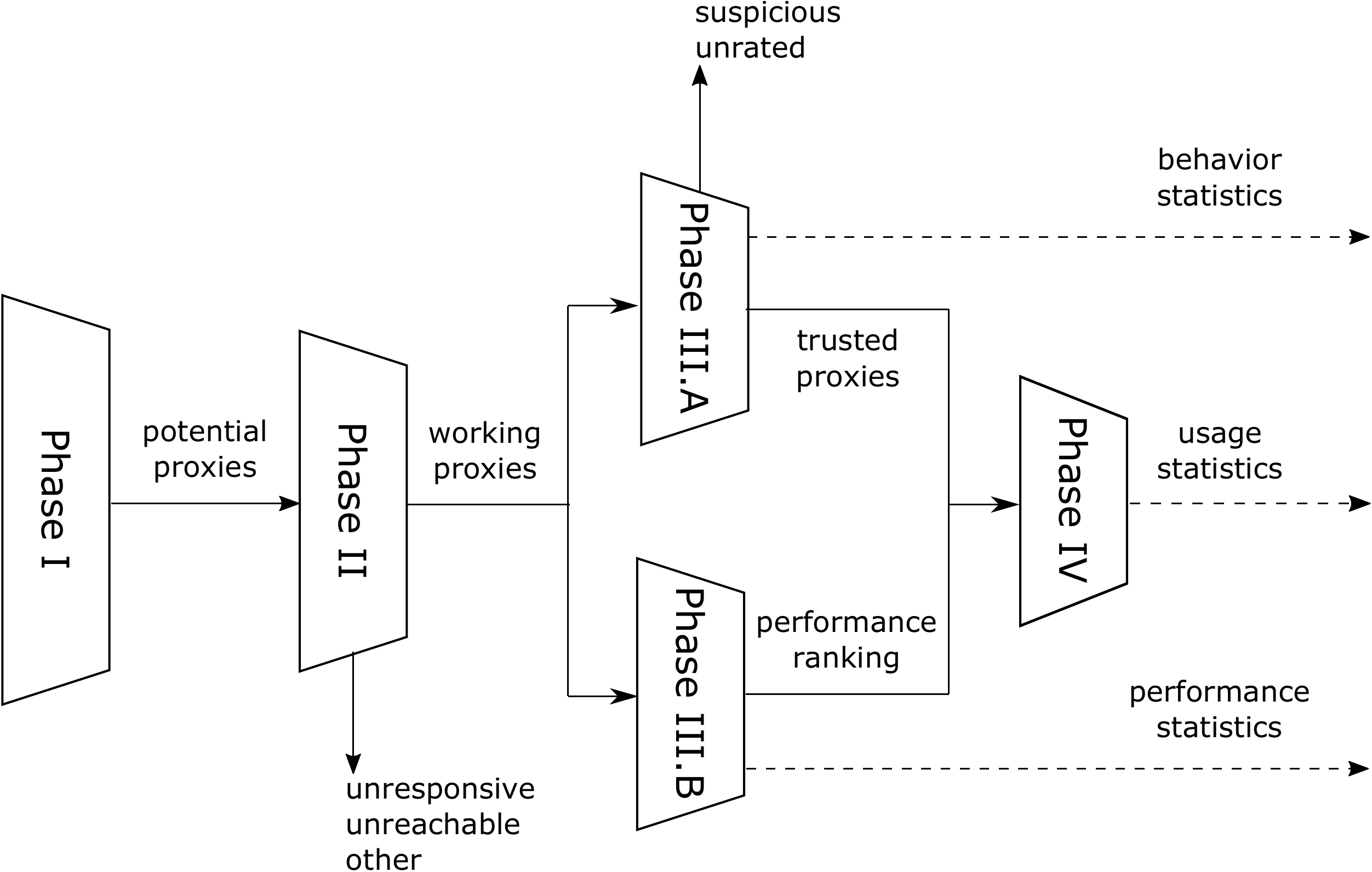, width=3.3in}
	\caption{\tool system overview.}
	\label{fig:overview}
\vspace{-.5cm}
\end{figure}

\tool shares some similarities with \texttt{proxycheck}~\cite{proxychecker}, a tool that can check the behavior of a proxy by using it to download few distinct objects hosted on a private webserver. Next, it labels a proxy as untrusted if the retrieved objects differ from the original ones even by a single bit, potentially generating a large number of false positives. Proxies are tested one at a time, which only allows to test $\sim$10,000 proxies a day. Despite some similarities, our approach is fundamentally different since we designed a funnel-shaped methodology (see Figure~\ref{fig:overview}) aiming to minimize false positives while maximizing performance, \eg scaling up to hundreds of thousands proxies per day.

\noindent\emph{In-path Manipulations.} A number of papers measure in-path web content manipulations by leveraging bait content served from a controlled host. Reis et al.,~\cite{reis08nsdi} focus on middleboxes and serve a page with an embedded JavaScript that detects and reports modifications. Their data-set contains 50,000 unique visits to their website, totaling 650 instances of content manipulation. Chung et al.,~\cite{ChungCM16} use the paid version of Hola~\cite{hola} (a peer-to-peer proxy network) to detect end-to-end violation in DNS, HTTP, and HTTPS traffic. They witness DNS hijacking, HTTP manipulations, image transcoding, and a few cases of TLS man-in-the-middle attempts. Many of the violations reported in~\cite{ChungCM16} are attributed  to ISPs and to (malicious) software running at Hola proxying peers.
Tyson et al.,~\cite{tyson17www} use the same approach to investigate HTTP header manipulations. They leverage Hola to gather 143k vantage points in 3,818 Autonomous Systems (ASes) and detect header manipulation in about 25\% of the ASes.
Weaver et al.,~\cite{WeaverKDP14} focus on transparent proxies using Netalyzr~\cite{KreibichWNP10} and detect HTTP proxy manipulations for 14\% of the connections.

Differently from all of the above, we look at performance and  content manipulations of free, non-transparent web proxies. We use bait content served from a controlled host, as well as real websites.
Our measurement platform also leverages real users by means of a plugin that provides easy proxy usage in exchange of anonymous statistics on how proxies are used in the wild.

\noindent\emph{Virtual Private Networks.}
Perta et al.,~\cite{PertaBTHM15} study privacy leaks in commercial VPN systems. Despite a VPN tunnel, they discover the following traffic leakages. First, IPv6 traffic is usually not tunneled. Second, poor management of the DNS configuration at the client may result in an adversary hijacking DNS requests and learning which websites a user visit. Similar issues are also reported by Ikram et al.,~\cite{IkramVSKP16} that analyze 283 Android VPN apps. The authors of~\cite{IkramVSKP16} also detect VPN apps with embedded tracking libraries and malware. Differently from these works, we focus on free web proxies that are a valid alternative to commercial VPNs in use-cases such as accessing geoblocked content. Apart from their behavior, we further assess their performance.



\begin{table*}[t]
\small
\centering
\begin{tabular}{rccccc}
\toprule

                 & \pbox{20cm} {\bf Phase I} & {\bf Phase II} & {\bf Phase III.A} & {\bf Phase III.B} & {\bf Phase IV}   \\
\midrule
{\bf \# clients} & 1 & 1 & 1 &$\sim$30 & up to 1,500\\
{\bf tools}      & beautifoulsoup & curl & PhantomJS/curl/OpenSSL & curl & Chrome plugin \\
{\bf main task}  & web-crawling & fetch 1KB synthetic object & fetch syntetic webpage& fetch real webpages & {interface with free proxies}\\
{\bf main goal}  & find potential proxies & find working proxies & test behavior & test performance & monitor performance and usage\\
{\bf frequency}  & daily & daily, on-demand & daily & every 5 minutes & user-controlled\\
{\bf classification} & potential  & \begin{tabular}{@{}c@{}}working/unresponsible/ \\ unreachable/other\end{tabular}    &  trusted/suspicious/unrated   & trusted/suspicious/unrated & ---\\
\\
\midrule
\end{tabular}
\caption{Key aspects of each phase in \tool.}
\label{tab:overview}
\vspace{-.5cm}
\end{table*}

\section{ProxyTorrent}
\label{sec:system}

This section describes \tool, a distributed measurement platform built to monitor the free proxy ecosystem.
Figure~\ref{fig:overview} shows an overview of the full system.

Due to the scale of the proxy ecosystem ---  potentially millions of machines~\cite{scott15ccc} --- we use a funnel-shaped methodology with several phases (see Figure~\ref{fig:overview}). Proxies are fed into the funnel and, at each phase, go through a series of tests of increasing complexity. Only proxies that pass a given phase are admitted to the next one. 
Since each phase decreases the number of proxies under test, we can progressively increase test complexity.
The last phase takes place at real proxy users, allowing to complement results of controlled experiments with measurements in the wild. Table~\ref{tab:overview} lists the key aspects of each phase.
In the remainder of this section, we describe all phases in detail.

%

\hfill \break
\noindent{\bf Phase I} discovers free proxies ($<$ip, port$>$ pairs) on the Internet by crawling several aggregator websites which regularly publish free proxy lists. 
Daily crawling runs from a single machine at our premises. The hosts discovered are used to populate  a list of ``potential proxies'' sorted by the last day when each proxy appeared on any of the webistes we crawl.

\hfill \break
\noindent{\bf Phase II} tests the potential proxies populated by Phase I for proxying capability.  We use curl~\cite{curl}, instrumented for full statistics and headers collection, to fetch a 1KB object via each potential proxy. The content is served using nginx~\cite{nginx} from a server hosted by Amazon (Ireland). Phase II runs daily from a single machine. It traverses the potential proxies list in order, and runs for up to 24 hours until either all proxies have been tested or time is over. This strategy rules out the least recently crawled potential proxies, in case the list becomes too big to be processed in a day. 

Each proxy is associated to a \emph{similarity score} computed as the ratio of common content across the webpage fetch performed with and without the proxy.  Accordingly, a similarity score of 1 means that the content fetched through the proxy is identical to one fetched without a proxy.




Phase II categorizes hosts as follows.
\emph{Unresponsive}: hosts for which either a connection or max duration timeout was triggered.\footnote{We measured empirically that 3 seconds (TCP handshake) and 30 seconds (maximum duration) are long enough for 95\% of the proxies.}
\emph{Unreachable}: hosts that either closed the TCP connection with a reset message or sent ICMP messages declaring the network or the requested host as unreachable.
\emph{Working}: hosts with a similarity score $\geq 0.5$, \ie that have correctly proxied at least 50\% of the 1KB object we serve. This threshold was empirically chosen to discard proxies returning errors or login pages, for which we empirically measured similarity scores lower than 0.3, on average. Note that proxies that largely alter a webpage might be caught in this rule as well. This is fine as far as finding safe working proxies, but it prevents the full behavioral analysis from phase III thus generating false negatives (see Section~\ref{sec:ecosystem:behavior}). We further classify working proxies as \emph{transparent}, \emph{anonymous}, or \emph{elite} (see Section~\ref{sec:back-related}) using HTTP headers collected both at the client and at the server. HTTP headers of all proxies are also analyzed to identify header manipulations that can be potentially malicious. Finally, Maxmind~\cite{maxmind} is also used to obtain country/AS information of each working proxy.
\emph{Other}: all remaining hosts that relay content substantially different from the expected one, \eg  all the hosts returning a login page (private or paid proxies) or an error page (misconfigured hosts).


\hfill \break
\noindent{\bf Phase III} tests working proxies with respect to behavior and performance. To assess a proxy behavior, we use the previous methodology of  comparing proxied content with content received when no proxy is used. Compared to Phase II, we introduce a headless browser, real content, HTTPS testing, and clients at multiple locations. For performance, we measure both \emph{page download time} (PDT) and \emph{page load time} (PLT). PDT is the time required to download the index page of a website; PLT is the time from when a browser starts fetching a website to the firing of the JavaScript \texttt{onLoad()} event, which occurs once the page's embedded resources have been downloaded, but possibly before all objects loaded via scripts are downloaded. Phase III consists of two parts (A and B) which both operate on the set of working proxies identified by Phase II within the last 7 days.

\hfill \break
\noindent{\bf Phase III.A} runs daily from a single machine at our premises. It uses PhantomJS~\cite{phantomjs}, a popular headless browser, to fetch a realistic website we serve. We designed the website to include elements that could trigger content manipulation by a proxy: a landing page \texttt{index.html} (83.7KB), two \texttt{javascripts} (635B and 22.9KB), two \texttt{png} images (1.5KB and 13.5KB), and a \texttt{favicon} (4.3KB). Our bait webpage is similar to the one set up by related work that looks for en-route content manipulation~\cite{ChungCM16}.

Data is collected as an HTTP Archive (HAR); for this, we have extended PhantomJS's HAR capturer\footnote{http://phantomjs.org/network-monitoring.html} to also dump the actual content downloaded. The HAR file includes detailed information about which object was loaded and when, as well as PLT. We stop PhantomJS either one second after the \texttt{onLoad()} event, to allow for potentially pending objects to be downloaded, or after a 45 second maximum duration timeout. Compared to Phase II, we increase the maximum duration timeout to account for an overall more complex operation.

Phase III.A also checks for issues with X.509 certificates. First, we use curl to connect to our server over port 433 and compare the X.509 certificate presented to the client with our original certificate (provided by LetsEncrypt~\cite{let_encrypt}). If curl detects any issue with the certificate, we use OpenSSL to download the X.509 certificates from our website as well as from two popular websites.\footnote{\url{https://www.theguardian.com} and \url{https://www.google.com}}

Phase III.A classifies a working proxy as \emph{trusted}, \emph{suspicious}, or \emph{unrated}. Trusted proxies serve the expected content with no alteration and do not replace or modify X.509 certificates. Suspicious proxies alter the relayed traffic, \eg by adding unsolicited content or by not relaying the expected X.509 certificates. Finally, \emph{unrated} proxies operate at such a slow speed that they are incapable to serve the full content requested within the maximum duration allowed. The partial content they serve is as expected, otherwise we mark them as suspicious. Phase III.A quantifies the performance of trusted and suspicious proxies using the PLT of our realistic website.

\hfill \break
\noindent{\bf Phase III.B} runs daily from 30 Planetlab nodes. Curl is used to fetch, via each proxy in the working proxy list,  the landing pages of Alexa's top websites. Precisely, we construct two 1,000-website lists from Alexa with support for HTTP and HTTPS, respectively. For each proxy and fetched page, Phase III.B reports both PDT and similarity score. Proxies are tested mostly against HTTP websites; only once every 10 tests a proxy is also tested for HTTPS support by fetching a random website from the HTTPS list. We empirically measured that phase III.B is currently capable of testing each working proxy at least once every 5 minutes.

\hfill \break
\noindent{\bf Phase IV} allows to both test free web proxies in the wild, as well as to learn how free proxies are used. It runs on the machines of the users that installed \plugin,\footnote{https://goo.gl/y86fOy} a Chrome plugin we developed to help users finding free proxies. Users pick the desired anonymity level (transparent, anonymous, elite) and location, and \plugin automatically sets up a free proxy based on input from \tool.  In order to minimize risk and maximize usability, we only consider proxies that have been labeled as \emph{trusted} in Phase III.A, and that have shown the best performance in Phase III.B. At any time the user can request a new proxy either to reflect a new preference or in case of failure.



\plugin reports statistics per \emph{download}, which captures all the events in a browser's tab transitioning from one URL to another, usually in response to directly typing a URL, refreshing or aborting the load of a webpage, clicking a link within a page, etc. We leverage Chrome's \texttt{webNavigation} APIs to identify the beginning and end of a download. For each download, the following statistics are collected: timestamps associated to the beginning and end of a download, PLT, number of requests per protocol type (HTTP/HTTPS), amount of bytes downloaded (HTTP/HTTPS),  navigation errors (if any). No personal information, such as IP address, browser/OS information, or URLs are reported at any time.

\begin{table}[t]
\small
\centering
\begin{tabular}{rccccc}
\toprule

           & \pbox{20cm} {\bf Total} & {\bf Unresp.} & {\bf Unreach.} & {\bf Other} & {\bf Working} \\
\midrule

{\bf Crawling}   & 0.16M   & 0.11M   & 0.04   & 8,000 & 2,895   \\
{\bf Zmap}       & 29.1M   & 17M     & 5.66M  & 6.4M. & 2,518   \\
\midrule
{\bf 8080}       & 13.5M   & 6.6M    & 2.2M   & 4.7M   & 376    \\
{\bf 8081}       & 7.2M    & 4.65M   & 1.55M  & 0.95M  & 171    \\
{\bf 8118}       & 4.7M    & 3.45M   & 1.15   & 0.1M   & 1,093  \\
{\bf 3128}       & 3.7M    & 2.29M   & 0.76M  & 0.65M  & 878    \\
\\
\midrule
\end{tabular}
\caption{Crawling and scanning (Zmap) summary, June 18th 2017. Results in the last four rows refer to scanning per port.}
\label{tab:summ_magn}
\vspace{-.5cm}
\end{table}

\section{The Free Proxy Ecosystem}
\label{sec:ecosystem}
This section characterizes the free proxy ecosystem. We first quantify its magnitude and evolution over time. Next, we  provide data supporting (or not) the preconception that free proxies are mostly malicious and tend to manipulate served content. We then conclude by assessing the ecosystem performance and by providing some evidence on how free proxies are used in the wild. We report on 10 months worth of data (January-October, 2017) spanning more than 180,000 proxies and 1,500 users.

\hfill \break
\noindent{\bf Limitations}
We acknowledge from the outset the limitations of our methodology. According to our findings, around 10\% of the working proxies every day exhibit malicious behavior by either injecting content, manipulating headers, or by replacing X.509 certificates. This is a lower bound to the fraction of malicious proxy since an exhaustive behavioral analysis by only controlling a few clients and servers is out of reach. We stress, however, that related work using a setup similar to ours, shares the same limitations~\cite{WeaverKDP14,ChungCM16,reis08nsdi,tyson17www}.

A proxy could behave maliciously only in some cases in order to avoid detection. For example, it may decide to manipulate content based on contextual factors, such as the client IP address, the domain requested, etc. Our experiments indicate that only 20\% of the malicious proxies manipulate the content of each requested page while many (40\%) do so only for one out of ten pages requested. Furthermore, there is no guarantee that a proxy that in our experiment proxied traffic without alterations, will not manipulate content when serving other users. Perhaps the content we requested or the IP address of our clients simply did not trigger content manipulation at the proxy. Another form of malicious behavior that we cannot fully assess is user tracking and profiling. Our experiments reveal several attempts to inject tracking/fingerprinting code, but we cannot rule out that even innocent-looking proxies carry out user profiling by simply leveraging the IP address of the user and her list of requests.
We nevertheless argue that \tool improves the current situation for proxy users that are clueless on whether a given proxy is performing any kind of malicious activity with the relayed traffic. Furthermore, \tool raises the bar for malicious proxies to avoid detection.


\subsection{Characterization}
\label{sec:results:avail}
\noindent{\bf Magnitude.} Table~\ref{tab:summ_magn} shows a snapshot of the free proxy ecosystem (June 18th, 2017). We chose this date since, at that time, we supplemented \tool's  crawling strategy by scanning the full IPv4 space and targeting the most popular proxy ports according to the aggregator websites. Our goal is to understand the \emph{coverage} of the aggregator websites we crawl. IPv4 address scanning leverages Zmap~\cite{zmap_2013} from a number of machines we control. Because of the ethical issues related to port-scanning, we run the scan only once. While we test the found proxies to categorize them,  we \emph{do not} use proxies found exclusively via scanning in the following experiments nor we make them available to \plugin users.

Table~\ref{tab:summ_magn} reports proxies obtained by \emph{crawling} the aggregator websites (first row), and the one found via port-scanning (second row). The table distinguishes between four hosts categories: unreachable, unresponsive, working, and other (see Section~\ref{sec:system}). Crawling yields a higher ratio of working proxies (2,895 out of approximately 160k) compared to port-scanning (2,518 out of more than 29M). Only 719 proxies appear in both data-sets. Regardless of the discovery strategy, the table shows that most hosts are either unresponsive or unreachable, and that only a few thousand hosts can actually be labeled as working proxies. The last four rows of Table~\ref{tab:summ_magn} show the breakdown of the proxies discovered via scanning by port.


\begin{figure}[t]
\centering
    \psfig{figure=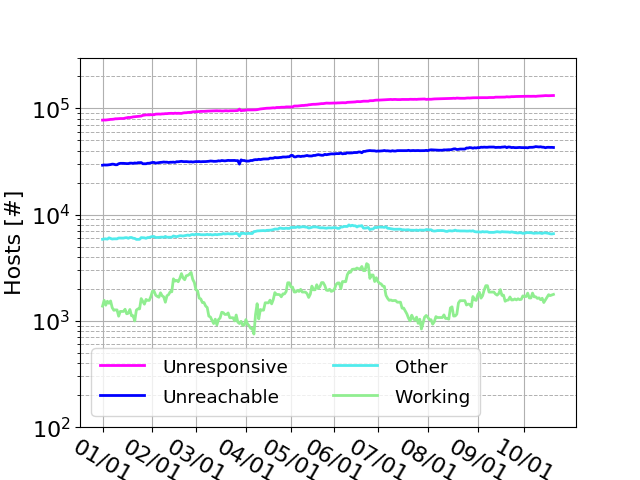, width=3in}	
	\caption{Time evolution of host classification: unreachable, unresponsive, working, and other.\label{fig:availability-a}}
\end{figure}

\begin{figure}[t]
\centering
    \psfig{figure=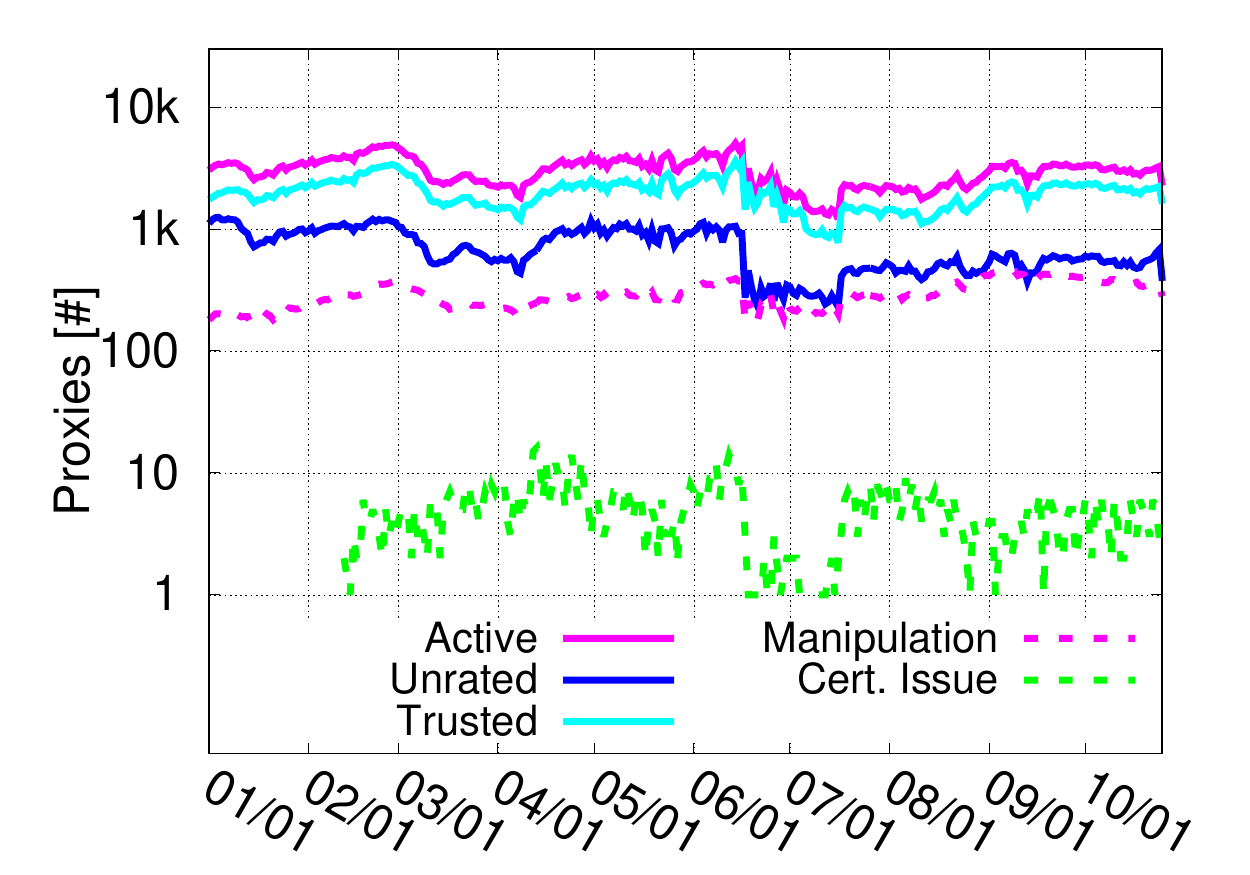, width=3in}
	\caption{Time evolution of working proxies by category: unrated, trusted, and  suspicious (either data or TLS certificate manipulation)\label{fig:availability-b}}
\end{figure}

Figure~\ref{fig:availability-a} shows the evolution over time of each proxy category as defined in Phase II. On the first day, we bootstrap \tool with a list of potential proxies containing 118,915 hosts ($<$ip, port$>$ pairs) collected on specialized forums. We then daily supplement such list via crawling. Overall, the figure shows that the working proxy category has a different trend than the others. While the number of hosts in each category increases over time, the number of working proxies oscillates between 900 and 3,000.


We now focus on the (small) core of working proxies for which further testing was conducted. Figure~\ref{fig:availability-b} shows the evolution over time of the \emph{active} proxies, \ie the set proxies that were reachable during phase III.A at least once within a day.  Figure~\ref{fig:availability-b} also shows the evolution of the categories \emph{trusted}, suspicious (split between proxies that manipulate TLS certificates---\emph{cert. issue}---and proxies that manipulate actual content---\emph{manipulation}), and \emph{unrated}. \footnote{The curve \emph{cert. issue} starts from mid February, when we added HTTPS support to \tool.}

According to Figure~\ref{fig:availability-b}, every day roughly 66\% of active proxies are marked as trustworthy, while around 24\% are marked as unrated. Suspicious proxies amount to 10\% of the active, where 100-300 proxies manipulate proxied content and only a handful of them is caught replacing X.509 certificates. On average, 40\% of the proxies support HTTPS. The drop observed in all curves at mid-June is caused by a partial failure of our system resources.

\hfill \break
\takeaway{The proxy ecosystem is characterized by a small and volatile core of proxies surrounded by a large and increasing set of non-proxy hosts that are erroneously announced on aggregator websites.}

\begin{figure}[t]
\centering
	\psfig{figure=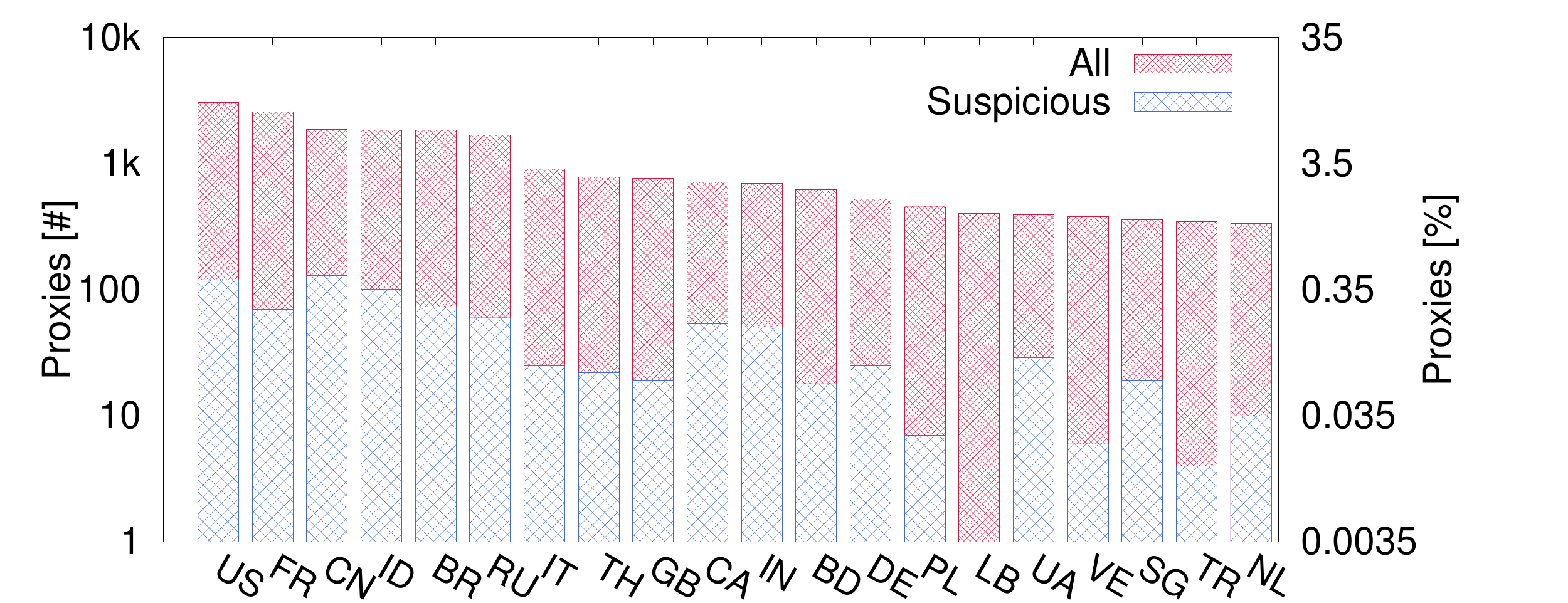, width=3in}

	\caption{Number of proxies per top 20 countries.}
	\label{fig:proxy-coutry}
\end{figure}

\begin{figure}
\centering
	\psfig{figure=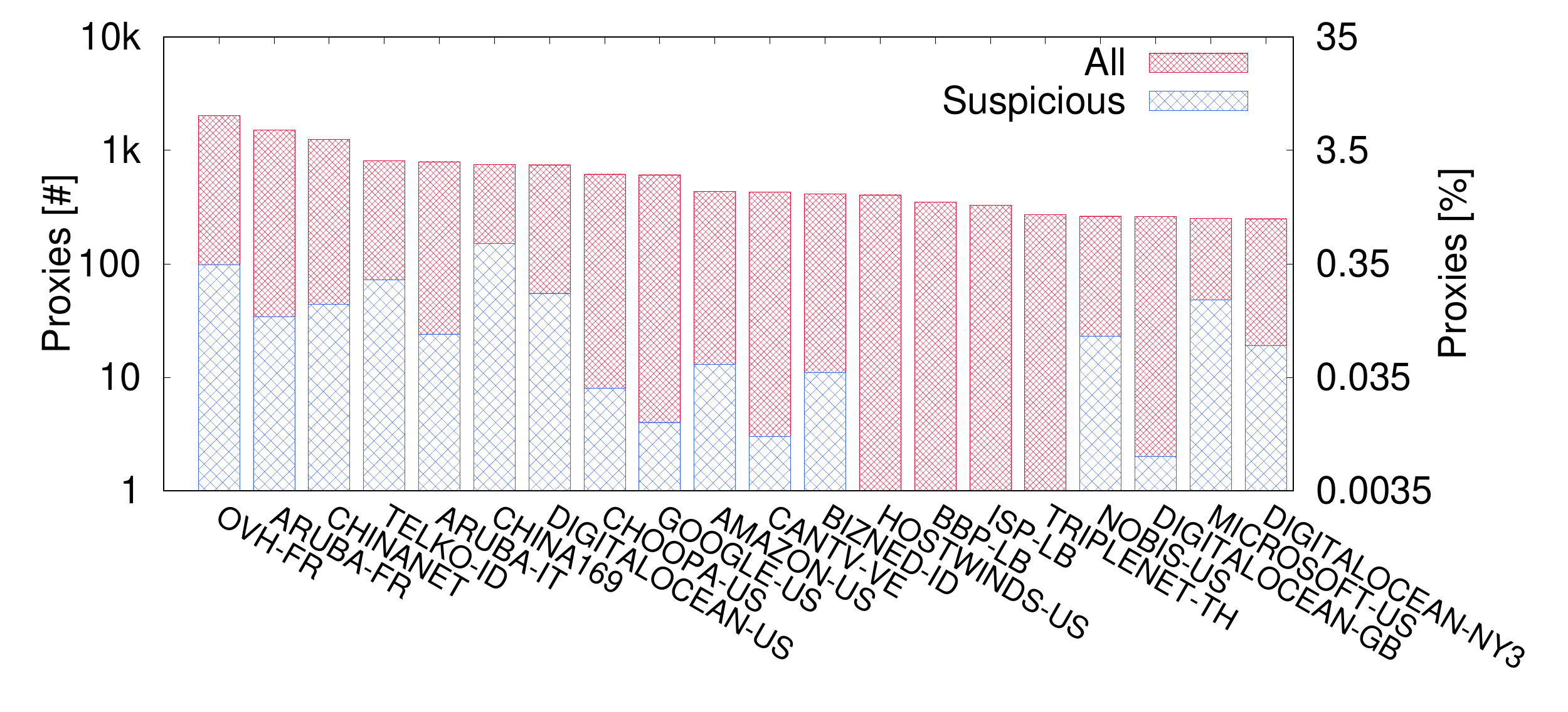, width=3in}
	\caption{Number of proxies per top 20 ASes.}
	\label{fig:proxy-as}
\end{figure}

\hfill \break
\noindent{\bf Geo-location.} Figure \ref{fig:proxy-coutry} and \ref{fig:proxy-as} show, for the top 20 countries and ASes, both the total number of proxies they host and the amount of suspicious ones. Both figures are computed considering all working proxies observed at least once during the six months monitoring period. USA (11\%), France (9\%) China (6.7\%), Indonesia (6.6\%),  Brazil (6.5\%),  and Russia (6\%) host 45\% of the proxies, while the remainder is scattered across 160 countries.  A similar distribution is observable for suspicious proxies, with the main differences being that China passes the US with respect to the number of suspicious proxies and the gap with France increases. As for the hosting ASs, about 28\% of proxies are concentrated in only six ASs, while the remaining proxies reside in 4,386 ASs. Both ISPs and cloud service providers appear in the top 20 ASs. 

\begin{figure}[t]
\centering
    \psfig{figure=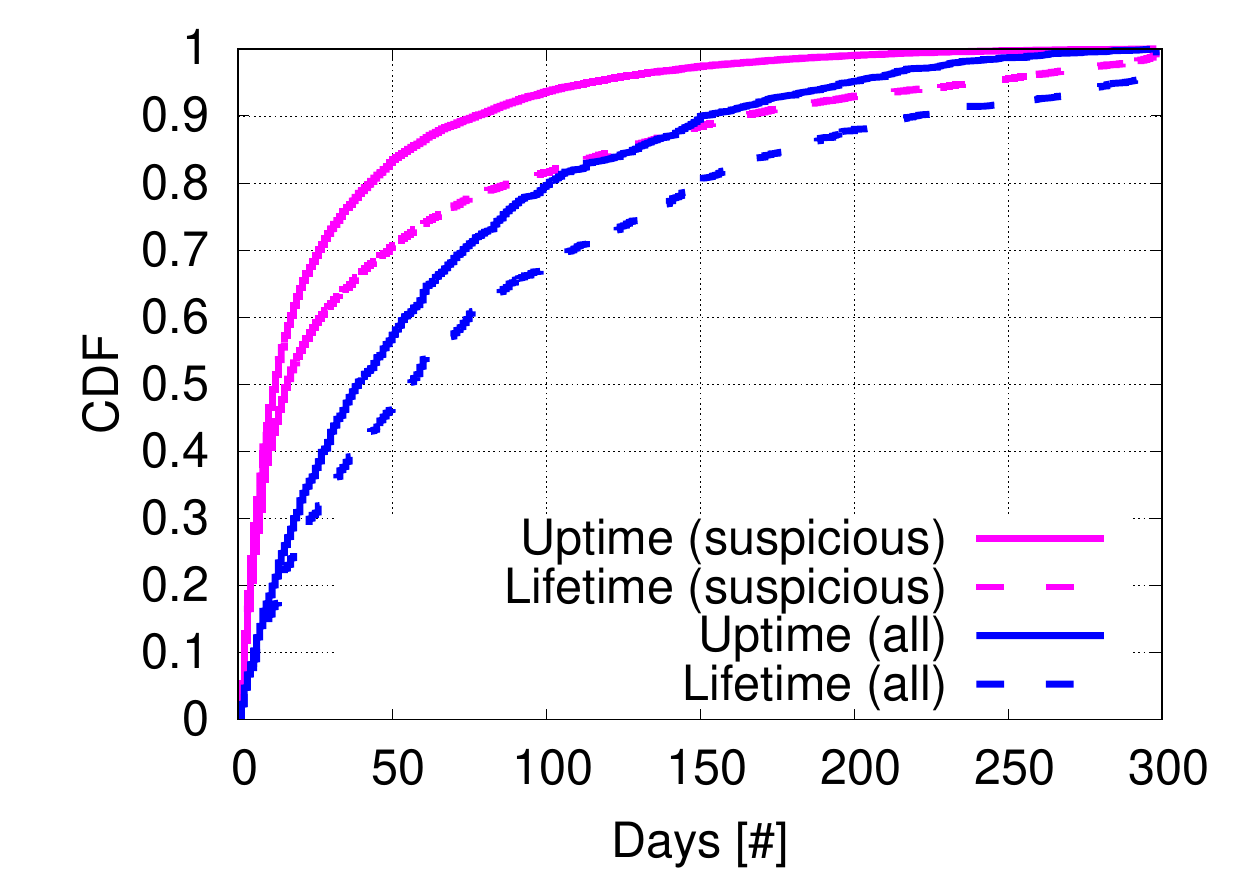, width=3in}
	\caption{CDF of lifetime and uptime for all proxies and the suspicious ones.\label{fig:availability-c}}
\end{figure}

\hfill \break
\noindent{\bf (In)stability.} Next, we explore the stability of the proxies located in the (usable) core of the free proxy ecosystem. We report their \emph{lifetime}, the number of days between the first and the last time a proxy has been active, and their \emph{uptime}, the number of days a proxy was active within its lifetime. Both metrics are derived using a proxy's IP address and port as an identifier; our estimates are thus lower bounds in presence of dynamic addressing. Figure~\ref{fig:availability-c} shows the CDF of lifetime and uptime over 10  months, distinguishing between all proxies and the suspicious ones. Proxies tend to have a long uptime, \eg 55\% of the proxies are available for their whole lifetime, regardless if they are suspicious or not. The figure also shows that suspicious proxies have a significantly shorter lifetime compared to the rest of the ecosystem, \eg a median lifetime of 15 versus 35 days. 

Roughly half of the monitored proxies lasts up to a month.
This result suggests that free proxies are fairly unstable over time. This can be due to dynamic addressing, for example when proxies run on residential hosts where they get their IP assigned by a dhcp server. Another possible reason is that some proxies serve public traffic due to misconfigurations that are eventually discovered and fixed by their administrator. The shorter lifetime measured for suspicious proxies could also be intentional, \ie frequent changes to the IP address might be used as a mean to circumvent banning from remote servers.

\hfill \break
\takeaway{The core of the free proxy ecosystem is characterized by an high level of instability which makes locating a usable proxy extremely challenging. Half of this core resides in a handful of countries, with the US leading the pack of trusted proxies and China the pack of suspicious ones.}

\subsection{Behavior}
\label{sec:ecosystem:behavior}
Differently from above, the following figures are aggregated statistics
over the 10 month monitoring period. We discovered 39,143 working proxies of which 16,700 (42\%) are classified as unrated, 1,833 (4.5\%) as suspicious and 20,610 (53.5\%) as trusted. Excluding unrated proxies---that do not serve enough content to enable a classification---8.2\% of proxies are  suspicious, and  91.8\% are trusted. This subsection focuses on suspicious proxies to comment on their behavior in detail.

\vspace{-0.1in}
\hfill \break
\noindent{\bf Suspicious Behavior Classification}
Content manipulated by suspicious proxies can be summarized as follows: \texttt{html} (74\% of all manipulated traffic), \texttt{javascripts} (24\%), and \texttt{images} (2\%). Unsolicited content injection mostly consists of \texttt{javascripts}, though we also spotted few \texttt{php} and image injections. Overall, we witnessed 228 unique manipulations of served content --- this implies that several proxies manipulate traffic in the same way. Also, suspicious proxies do not manipulate traffic at each request: only 20\% of them manipulate traffic all the time, while 40\% do it less than 10\% of the time.

In order to better understand the purpose of content manipulation, we resort to visual inspection. To minimize the effort, we first cluster manipulated content using affinity propagation clustering~\cite{frey2007clustering}. Specifically, we consider each piece of altered or injected content as a string and compute the distance matrix required by the clustering algorithm using the edit distance between each pair of strings. 

Among the output clusters, two of them cover about 60\% of the content manipulation instances. The first cluster contains 84 instances of \emph{ad injection code}, of which 50 can be linked to two companies that provide hotspot monetization services. The second cluster contains 47 instances of \emph{fingerprinting/tracking code}, mostly javascript code that attempts to identify a user. Thirty out those 47 instances include \texttt{rum.js}, a popular library to monitor user-webpage interactions. Although \texttt{rum.js} is commonly used by CDN providers, there is no apparent motivation for a free proxy to inject such code.

The remaining clusters include the following instances of injected code. Nine instances, imputable to only two proxies, display \emph{religious-related content}. Four times we witness metadata of \emph{pyweb}, a popular proxy rewriting tool for live web content. Pyweb's metadata triggerered our detection, but further inspection shows no actual content rewriting. Finally, we could not figure out the semantics of the remaining 84 content manipulations either because they were obfuscated or because they were only a few bytes in size.

\hfill \break
\takeaway{Few content manipulation strategies exist that are shared among many proxies, advertisement injection being the most frequent one. Suspicious proxies do not manipulate traffic constantly; \tool's continuous monitoring is thus paramount to detect such proxies.}
\hfill \break

\noindent{\bf Invalid X.509 Certificates}
HTTPS is supported by 17,350 proxies (about 44\% of the working proxies) and 0.9\% of them (173 proxies) were caught interfering with TLS handshakes.
The most common behavior among such proxies is to replace the original certificate with a self-signed one showing vague \texttt{CommonName} attributes such as ``https'' or ``US''. Three proxies provide certificates with \texttt{CommonName} matching the original domain but signed by ``Zecurion Zgate Web'', a company offering corporate gateways to mitigate information exfiltration, and ``Olofeo.com'', a French company that offers managed security services. 
Only one proxy delivers a certificate chain of size two, where the leaf certificate has the expected \texttt{CommonName} but the root certificate has \texttt{CommonName} set to ``STATESTATESTATESTATESTATE'').  The issuer of this certificate is \texttt{wscert.com}, a domain expired as of February 2017.

\hfill \break
\takeaway{Attempts of TLS interception are rare in the free proxy ecosystem. Modern browsers would easily detect these potential attacks and inform the user. Yet previous work has shown that users tend to click through warnings~\cite{akhawe13sec}.}
\hfill \break

\begin{figure}[t]
\centering
	\psfig{figure=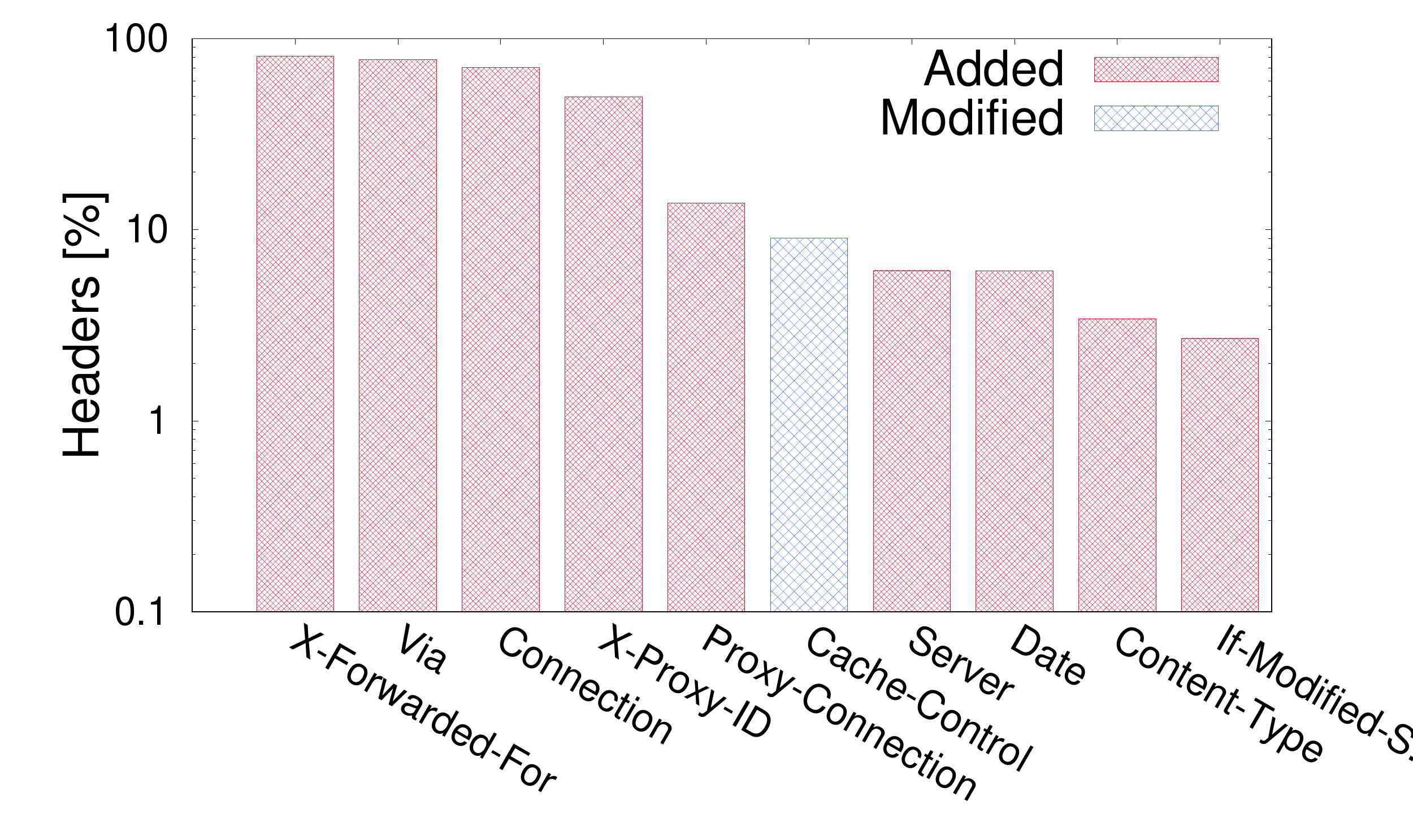, width=3in}
	\caption{Header manipulation: request headers.}
	\label{fig:eval-header-req}
\end{figure}

\begin{figure}[t]
\centering
	\psfig{figure=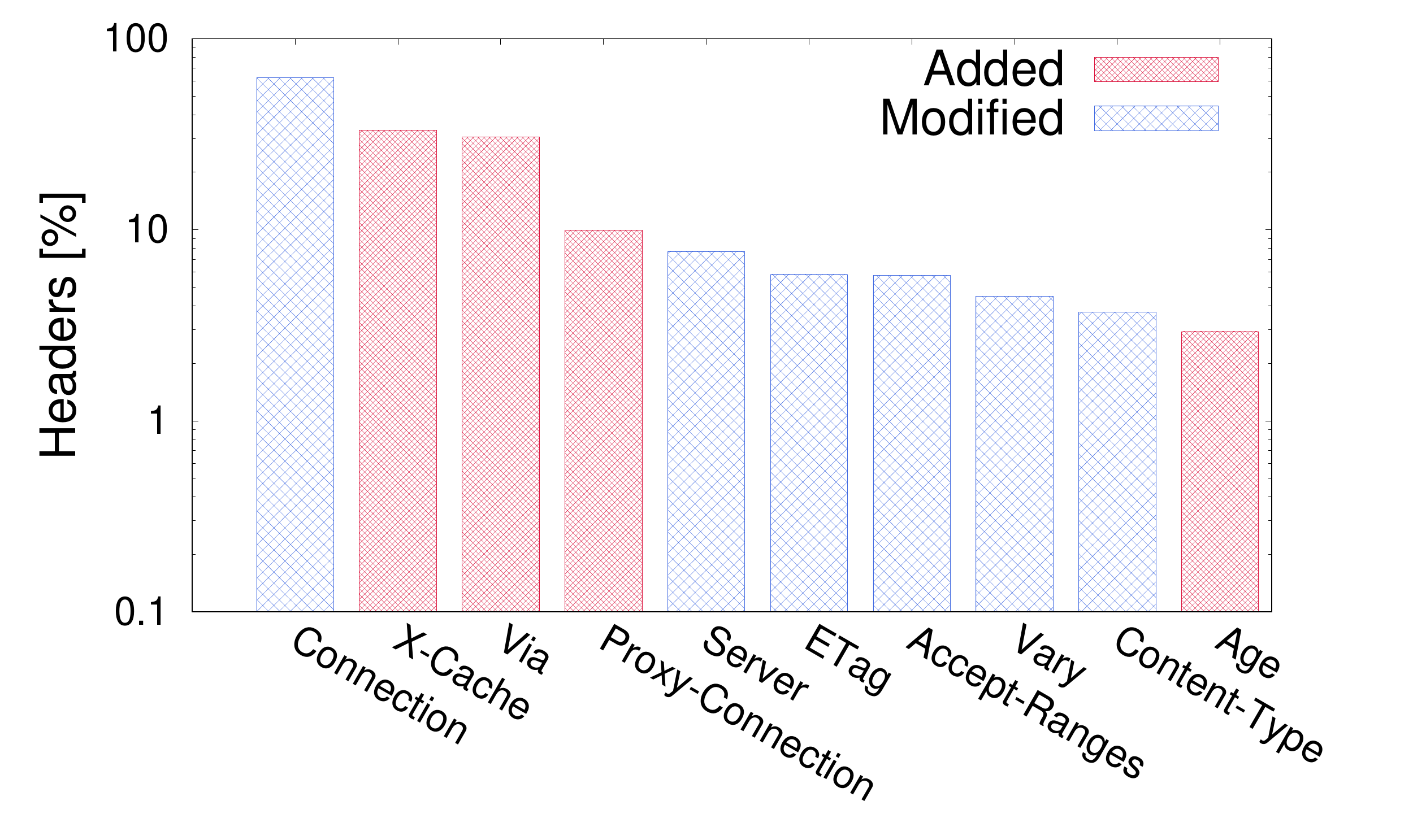, width=3in}
	\caption{Header manipulation: response headers.}
	\label{fig:eval-header-resp}
\end{figure}

\noindent{\bf Header Analysis} We now analyze HTTP request and response headers with the two-fold objective of understanding the level of anonymity provided by proxies, and if header manipulations by free proxies goes beyond traffic anonymization. First, we focus on the working proxies observed at least once during six months. Then, we extend our analysis to proxies categorized as \textit{other}, \ie proxies that relay a webpage that differs more than 50\% from our bait webpage (see Phase II in Section~\ref{sec:system}).

Figure~\ref{fig:eval-header-req} shows the top 10 request header modifications and injections observed. \texttt{Via}, \texttt{X-Proxy-ID}, \texttt{X-Forwarded-For}, and \texttt{Connection} are the most frequently added  headers. The first two headers are used by proxies to announce themselves to origin servers, while the third one specifies the client IP address to the origin server, when the proxy acts transparently. By leveraging those headers we classify proxies as: 1) transparent (77\%), proxies that reveal the original client IP to the server; 2) anonymous (6\%), proxies that preserve client anonymity but reveal their presence to the server; 3) elite (17\%), proxies that preserve client anonymity and do not announce themselves to the origin server.

\texttt{Connection} is another frequently injected header. Roughly 60\% of the proxies tested set it to \texttt{close} or \texttt{keep-alive}. This behavior is not surprising as this header is reserved for point-to-point communication, \ie between client and proxy or between server and proxy. The \texttt{Proxy-Connection} header plays a similar role, and it is also added in about 10\% of cases. \texttt{Cache-Control} is the only request header which is altered; about 10\% of proxies modify this header to accept cached content with a given \texttt{max-age} value, despite our testing tools explicitly specify not to serve cached content. We also observe that less than 1\% of proxies (not shown in Figure~\ref{fig:eval-header-req}) modify the user-agent by either removing it or specifying their own agents. While the exposure of the client user-agent reduces anonymity, it allows the server to optimize the content served based on the user device and application.

Figure~\ref{fig:eval-header-resp} shows the top 10 response header modifications and injections performed by working proxies. As for the request headers, the \texttt{Via} header is among the most frequently injected one; this is used by proxies to to announce themselves and their protocol capabilities to clients. About 30\% of proxies also add the \texttt{X-Cache} header to specify if the requested content was served from the proxy's cache or if a previously cached response is available. The most frequently modified header is the \texttt{Connection} header, that is either removed (50\% of cases) or set to \texttt{close}. As previously stated,  this is a common behavior as this header is connection specific and does not need to be propagated to the client. Finally, less than 10\% of the proxies modify the \texttt{Server} header to reflect the software they use, rather than the one of the origin server.

\begin{figure}[t]
\centering
	\psfig{figure=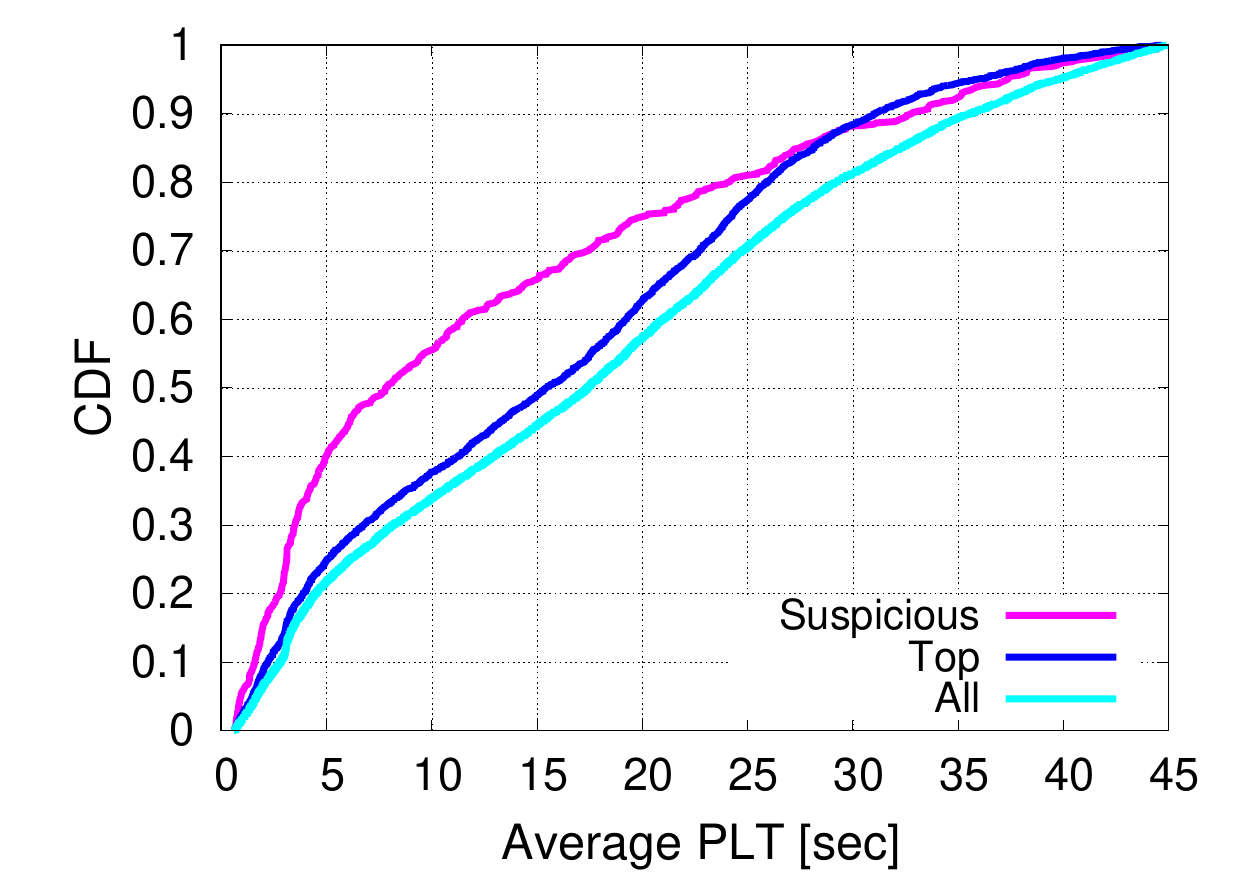, width=3in}
	\caption{CDF of average PLT per proxy distinguishing between suspicious, all, and best performing proxies.}
	\label{fig:proxy-performance-mix}
\end{figure}

\begin{figure*}[t]
\centering
\subfigure[Fraction of queries for the top 10 countries based on user preferences.]{
\psfig{figure=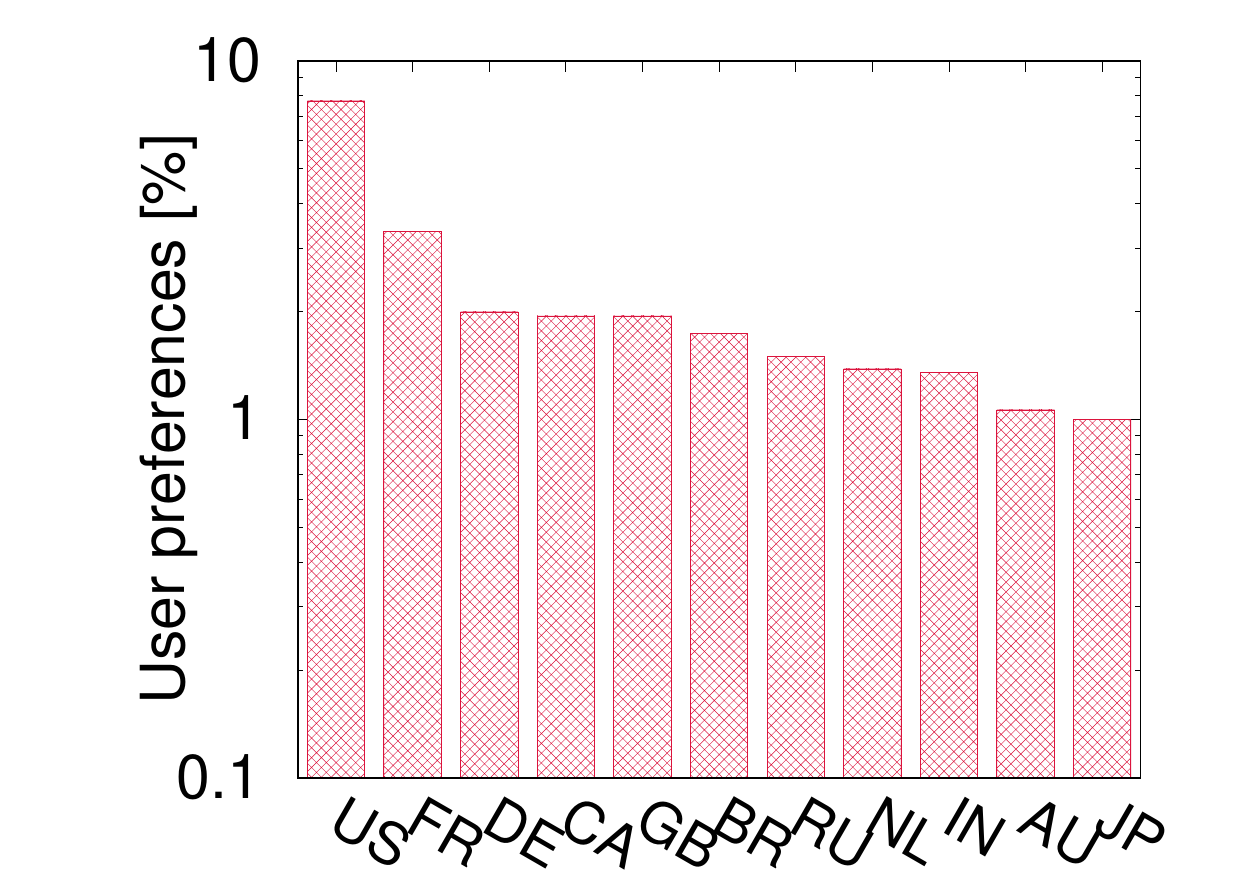, width=2.2in}\label{fig:res-dplane-queries}
}
\subfigure[Fraction of downloads and bytes proxied by the top 10 countries. ]{
	\psfig{figure=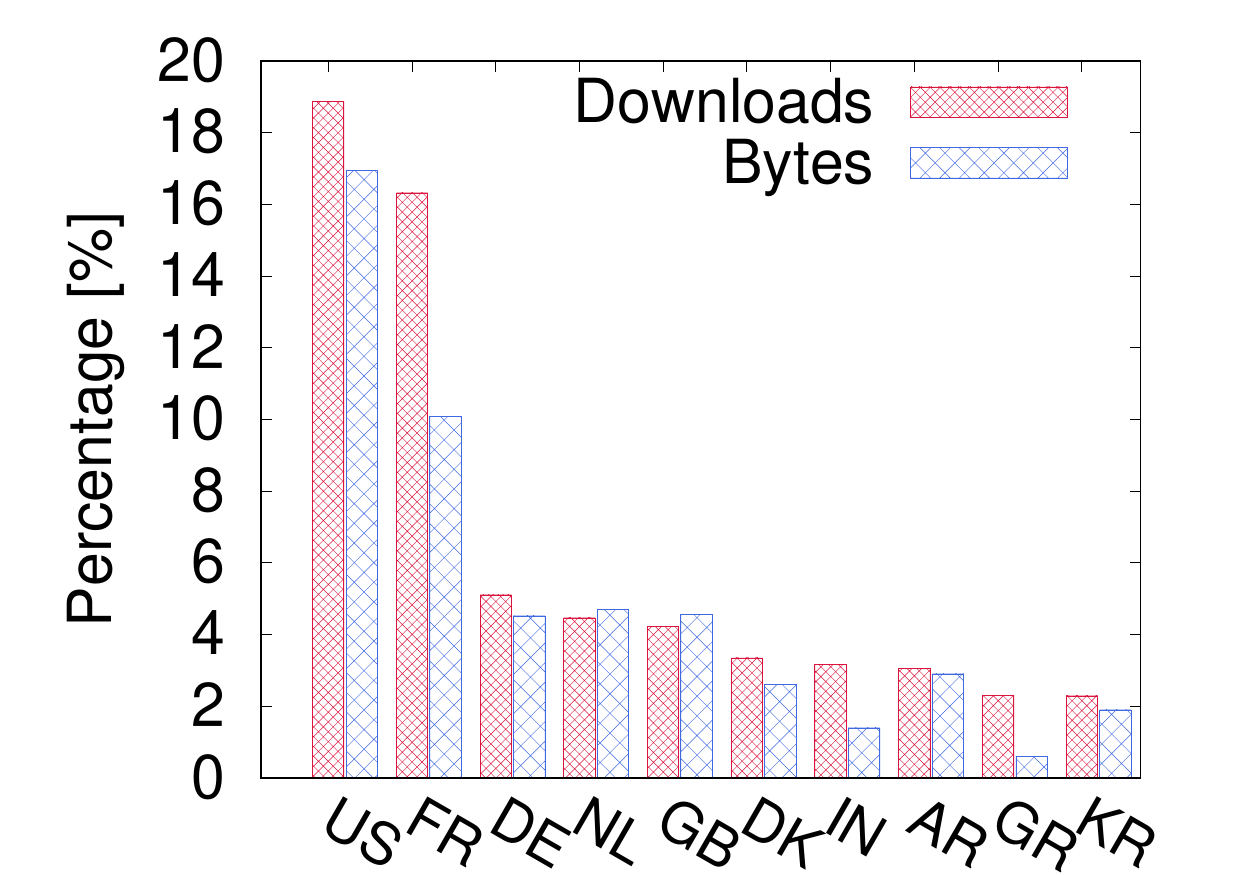, width=2.2in}\label{fig:res-dplane-countries}
	}
\subfigure[CDF of the percentage of geo-localized traffic.]{
	\psfig{figure=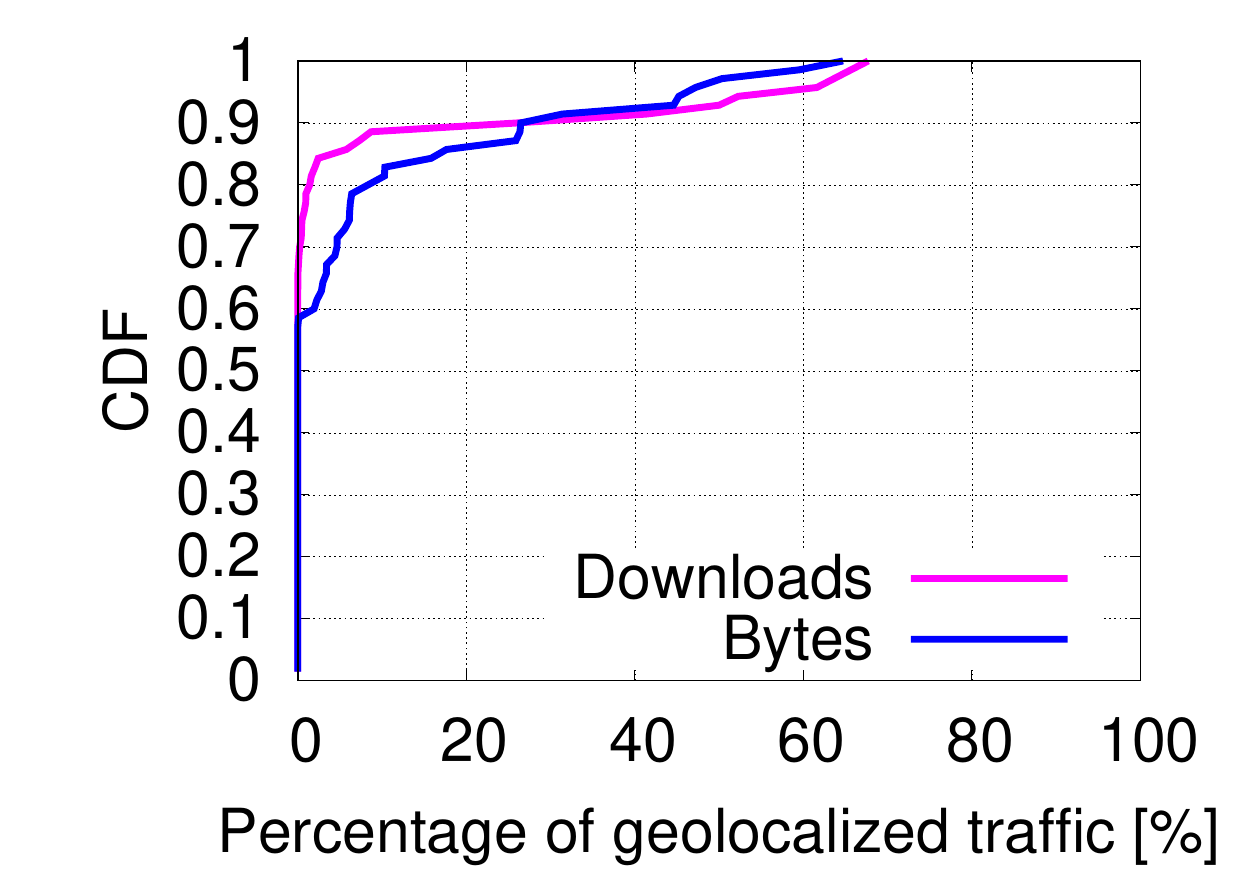, width=2.2in}\label{fig:res-dplane-geolocalization}
	}
	\caption{Proxy usage and geo-location analysis.}
\end{figure*}

We now focus on proxies categorized as \emph{others}. Similar observations as above hold; in addition, we observe a non negligible amount of \texttt{Set-Cookie} (5\%), \texttt{Access-Control-Allow-*} (1\%), and  \texttt{X-Adblock-Key} (0.5\%) headers in response to clients. The \texttt{Set-Cookie} header pushes a cookie to the client that may be used for tracking. The \texttt{Access-Control-Allow-*} headers are used to grant permission to clients to access resources from a different origin domain than the one currently in use. Both headers expose clients to malicious or unintended activities; however, they are also frequent for private and enterprise proxies. Because similar headers were not observed, at this scale, for the working proxies, we conclude that this behavior is unlikely malicious. Conversely, the \texttt{X-Adblock-Key} response header allows ads to be displayed at clients bypassing ad-blocker tools.  Proxies injecting this header likely return a modified version of our ``bait'' webpage including extra advertisements, which largely departs from the original page (similarity score $< 0.5$). Proxies categorized as \emph{others} were between 30,000 and 40,000; this analysis suggests that the similarity score rule introduces about 150-200 false negatives or about 1\%.




\hfill \break
\takeaway{HTTP header analysis reveals that the free proxy ecosystem is mostly composed of transparent proxies which announce themselves and/or reveal the client's IP address to the origin server. Suspicious header manipulation is rare; when present, it aims at ensuring that injected advertisements are not filtered by ad-blockers}.

\subsection{Performance}
We now investigate the \emph{performance} of the free proxy ecosystem, or how fast can free proxies deliver content to their users. We use page load time (PLT) as a performance metric since it accurately quantifies end user experience~\cite{eyeorg}. However, PLT also depends on the composition of a webpage, \ie its overall size and complexity in terms of number of objects. Accordingly, it has to be noted that PLT values from experiments in Phase III refer to our synthetic webpage---small size and only few objects---while PLT values for experiments in Phase IV refer to proxies usage in the wild, \ie overall bigger webpages with hundreds of embedded objects.

Figure \ref{fig:proxy-performance-mix} shows the CDF of the average PLT measured through each proxy, distinguishing between suspicious proxies (\emph{suspicious}), all proxies in the ecosystem (\emph{working}), and the best performing proxies \tool offers to its users via \plugin (\emph{top}). PLT values for both all and suspicious proxies are measured in Phase III, while PLT values for top proxies are measured in Phase IV. Failed downloads, where no PLT was measured are not taken into account.

Figure \ref{fig:proxy-performance-mix} shows that suspicious proxies are faster than other proxies in the ecosystem, \eg the median PLT they provide is 2.5x faster (7 seconds versus 18). The figure also shows that \tool correctly identifies the best performing proxies since their PLT measured at the user is 15\% faster than the rest of the ecosystem.

\hfill \break
\takeaway{Suspicious proxies are, on average, twice as fast as safe proxies. Faster connectivity may be used by malicious proxy as a bait to attract more potential victims. Our finding support the popular belief that free proxies are ``free for a reason''.}

\subsection{Usage}

We released \plugin---our Chrome plugin to facilitate discovery and usage of free proxies---on the Chrome Web store on March 17th 2017, and announced it via email, social media, and few forums on free proxies, anonymity, censorship circumvention, etc. At the time of writing \plugin has been installed by more than 1,500 users who generated about 1,3 Millions downloads, totaling 2~TBytes of HTTP/HTTPS traffic (1.5/0.5~TBytes, respectively).

We start by investigating user preferences in terms of both proxy location and anonymity level.  While for 70\% of the queries the users did  specify a country preference, they requested a specific anonymity level only for 16\% of their queries. This indicates that users are overall more interested in the proxy location than its anonymity level. According to user preferences, anonymity levels can be ranked as follows: transparent proxies (7\%), elite (5\%),  and anonymous (4\%). With respect to proxy locations, only 20\% of the queries are concentrated in the 10 most popular locations (see Figure~\ref{fig:res-dplane-queries}), while the remainder 80\% are spread across 120 countries. These top 10 countries are also among the ones where most proxies are located (see Figure~\ref{fig:proxy-coutry}). Since \plugin shows how many proxies are available per country, it is possible that user preferences have been influenced by this information.

Next, we investigate \emph{where} most of the traffic is proxied. Figure~\ref{fig:res-dplane-countries} shows the fraction of downloads and bytes for the top 10 countries only considering downloads where a country preference was set (70\% of the time). In this case, the distribution is heavily skewed towards the top 10 locations accounting, overall, for about 60\% of all downloads and bytes transferred. However, the ranking between the two figures  is  fairly similar.

Figure~\ref{fig:res-dplane-geolocalization} shows the distribution of ``geo-localized'' \plugin traffic, \ie traffic associated with a webpage hosted in the same country of the proxy being used. For this analysis, we temporarily (one month) extended the statistics collected by \plugin (see Section~\ref{sec:system}, phase IV) with a boolean value indicating whether the location of a requested website is the same of the proxy used. Website location is inferred by \plugin via website top level domain, if it is representative of a country, or via third party services providing website-to-location mapping.\footnote{\url{http://ip-api.com}, \url{http://www.ip-tracker.org/domain-to-location.php}} \plugin users have been informed of this data collection via a message shown on the proxy selection interface.

Figure~\ref{fig:res-dplane-geolocalization} shows that, on average, the websites accessed via a free proxy and the proxy itself are  hosted in the same country for 30\% of downloads and 20\% of bytes. Further, we observe no geo-localized traffic for half of the countries.  This results suggests that geo-blocking avoidance is not a prominent use-case for free web proxies. However, the figure also shows some countries with high percentages of geo-localized traffic, \eg 60\% in the US.


The geo-localized traffic observed in Figure~\ref{fig:res-dplane-geolocalization} could be attempts to access popular geo-blocked services like Hulu or Netflix in the US. In absence of URL visibility, we investigate whether these users are particularly concerned about leaking their IP/location, \ie are more likely to request anonymous and elite proxies. We find that for these downloads anonymous proxies are the most popular choice (70\%)---while normally being the least popular choice---followed by transparent (23\%) and elite (7\%).  Even if elite proxies provide a higher anonymity level than anonymous ones, they are less likely used to access geo-blocked content. This may be due to their name that does not clearly highlight strong anonymity to non-expert users, differently from \emph{anonymous} proxies.

\begin{figure}[t]
\centering
	\psfig{figure=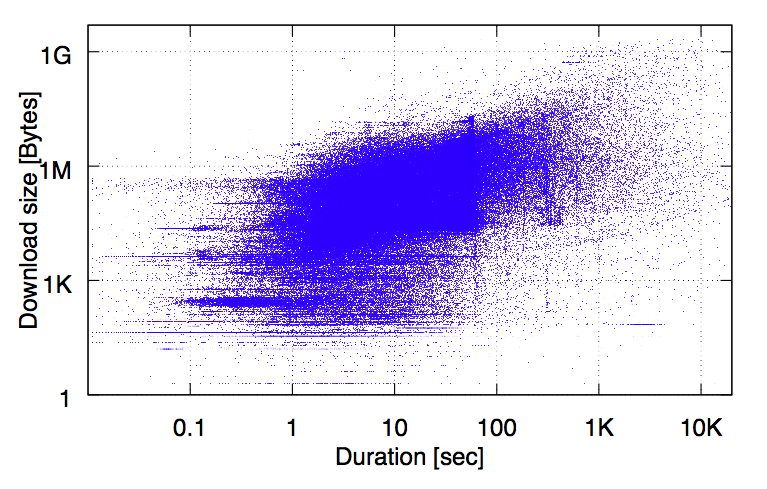, width=3in}
	\caption{Scatterplot of download size and duration.\label{fig:res-dplane-scatter}}
\end{figure}

Finally, we investigate which type of content is downloaded when using free web proxies. Our analysis relies on the little information \plugin collects to preserve its users privacy, \ie download size and duration. Figure~\ref{fig:res-dplane-scatter} shows a scatterplot of the size of each download (bytes) as a function of its duration (seconds). 95\% of downloads are short ($<$ 1 minute) and contain, on average, 500~KBytes. Even though  500~KBytes is less than the size of an average webpage --- \textit{httparchive}\footnote{\url{http://httparchive.org/}} reports it to be 2.9 MBytes --- these downloads relate to regular web browsing. The smaller download size we observe is due to: 1) \textit{httparchive} derives its statistics from crawling Alexa's top webpages while our workload is driven by real users that may visit a different set of websites, 2) our download size estimation is a lower bound on the actual webpage size as \plugin is oblivious to data retrieved from the browser's cache. Figure~\ref{fig:res-dplane-scatter} also shows a non-negligible amount of downloads lasting several minutes (0.1\%) and containing few 100~MBytes, as well as two very long downloads (up to couple of hours) containing few GBytes. These large downloads could be due to software or video downloads, live streaming, etc. We speculate the latter since no additional browsing activity was observed during these long sessions, \ie the user did not perform any other download suggesting that she could be watching the content being retrieved.

\hfill \break
\takeaway{\plugin has proven to be a valuable tool to shed some lights on how free proxies are used. By analyzing 2 TBytes of traffic generated by 1,500 users over 7 months, we identify web browsing as the most prominent user activity. Overall, geo-blocking avoidance is not a prominent use-case for free web proxies, with exception of countries hosting a lot of geo-blocked content like the US.}

Fueled by an increasing need of anonymity and censorship circumvention, the (free) web proxy \emph{ecosystem} has been growing wild in the last decade. Such ecosystem consists, potentially, of millions of hosts, whose reachability and performance information are scattered across multiple forums and websites. Studying this ecosystem is hard because of its large
scale, and because it involves two players out of reach: free proxies and their users. The key contributions of this work are \tool, a distributed measurement platform for the free proxy ecosystem, and an analysis of 10 months of data spanning up to 180,000 free proxies and 1,500 users. \tool leverages a funnel-based testing methodology to actively monitor hundreds of thousand free proxies every day. Further, it leverages free proxies users to understand how proxies perform and how they are used in the wild. The latter is achieved via a Chrome plugin we developed which simplifies the hard task of finding a \emph{working} and \emph{safe} free proxy in exchange of anonymous proxy usage statistics. Our analysis shows that the free proxy ecosystem consists of a very small and volatile core, less than 2\% of all announced proxies with a lifetime of few days. Only half of the proxies in this core have good enough performance to be used. However, users should be aware that about 10\% of the best working proxies are ``free for a reason'': ads injection and TLS interception are two examples of malicious behavior we observed from such proxies.  Finally, the analysis of more than 2~Terabytes of proxied traffic shows that free proxies are mostly used for web browsing and that geo-blocking avoidance is not a prominent use-case.

\newpage
\bibliographystyle{ACM-Reference-Format}
\bibliography{biblio-new}

\end{document}